\documentclass{article}
\usepackage{spconf,amsmath,amsfonts,graphicx}
\usepackage{amssymb,textcomp,mathtools}
\usepackage{bm,upgreek,algorithm,hyperref}
\RequirePackage{doi}

\def\thline{\noalign{\hrule height 1.0pt}}

\renewcommand{\vec}[1]{\bm{\mathrm{#1}}}

\title{TasNet: time-domain audio separation network for real-time, single-channel speech separation}
\name{Yi Luo \qquad Nima Mesgarani}

\address{Department of Electrical Engineering, Columbia University, New York, NY}

\begin{document}
\ninept
\maketitle
\begin{abstract}
 Robust speech processing in multi-talker environments requires effective speech separation. Recent deep learning systems have made significant progress toward solving this problem, yet it remains challenging particularly in real-time, short latency applications. Most methods attempt to construct a mask for each source in  time-frequency representation of the mixture signal which is not necessarily an optimal representation for speech separation. In addition, time-frequency decomposition results in inherent problems such as phase/magnitude decoupling and long time window which is required to achieve sufficient frequency resolution. We propose Time-domain Audio Separation Network (TasNet) to overcome these limitations. We directly model the signal in the time-domain using an encoder-decoder framework and perform the source separation on nonnegative encoder outputs. This method removes the frequency decomposition step and reduces the separation problem to estimation of source masks on encoder outputs which is then synthesized by the decoder. Our system outperforms the current state-of-the-art causal and noncausal speech separation algorithms, reduces the computational cost of speech separation, and significantly reduces the minimum required latency of the output. This makes TasNet suitable for applications where  low-power, real-time implementation is desirable such as in hearable and telecommunication devices. 
\end{abstract}
\section{Introduction}
\label{sec:intro}

Real-world speech communication often takes place in crowded, multi-talker environments. A speech processing system that is designed to operate in such conditions needs the ability to separate speech of different talkers. This task which is effortless for humans has proven very difficult to model in machines. In recent years, deep learning approaches have significantly advanced the state of this problem compared to traditional methods \cite{huang2015joint, zhang2016deep, isik2016single, kolbaek2017multitalker, chen2017deep, luo2017speaker}.  

A typical neural network speech separation algorithm starts with calculating the short-time Fourier transform (STFT) to create a time-frequency (T-F) representation of the mixture sound.  The T-F bins that correspond to each source are then separated, and are used to synthesize the source waveforms using inverse STFT. Several issues arise in this framework. First, it is unclear whether Fourier decomposition is the optimal transformation of the signal for speech separation. Second, because STFT transforms the signal into a complex domain, the separation algorithm needs to deal with both magnitude and the phase of the signal. Because of the difficulty in modifying the phase, the majority of proposed methods only modify the magnitude of the STFT by calculating a time-frequency mask for each source, and synthesize using the masked magnitude spectrogram with the original phase of the mixture. This imposes an upper bound on separation performance. Even though several systems have been proposed to use the phase information to design the masks, such as the phase-sensitive mask \cite{erdogan2015phase} and complex ratio mask \cite{williamson2016complex}, the upper bound still exists since the reconstruction is not exact. Moreover, effective speech separation in STFT domain requires high frequency resolution which results in relatively large time window length, which is typically more than 32 ms for speech \cite{isik2016single, kolbaek2017multitalker, chen2017deep} and more than 90 ms for music separation \cite{luo2017deep}. Because the minimum latency of the system is bounded by the length of the STFT time window, this limits the use of such systems when very short latency is required, such as in telecommunication systems or hearable devices. 

A natural way to overcome these obstacles is to directly model the signal in the time-domain. In recent years, this approach has been successfully applied in tasks such as speech recognition, speech synthesis and speech enhancement \cite{sainath2015learning, ghahremani2016acoustic, oord2016wavenet, mehri2016samplernn, pascual2017segan}, but waveform-level speech separation with deep learning has not been investigated yet. In this paper, we propose Time-domain Audio Separation Network (TasNet), a neural network that directly models the mixture waveform using an encoder-decoder framework, and performs the separation on the output of the encoder. In this framework, the mixture waveform is represented by a nonnegative weighted sum of $N$ basis signals, where the weights are the outputs of the encoder, and the basis signals are the filters of the decoder. The separation is done by estimating the weights that correspond to each source from the mixture weight. Because the weights are nonnegative, the estimation of source weights can be formulated as finding the masks which indicate the contribution of each source to the mixture weight, similar to the T-F masks that are used in STFT systems. The source waveforms are then reconstructed using the learned decoder. 

This signal factorization technique shares the motivation behind independent component analysis (ICA) with nonnegative mixing matrix \cite{wang2010nonnegative} and semi-nonnegative matrix factorization (semi-NMF) \cite{ding2010convex}. However unlike ICA or semi-NMF, the weights and the basis signals are learned in a nonnegative autoencoder framework \cite{hosseini2016deep, lemme2012online, chorowski2015learning, smaragdis2017neural}, where the encoder is a 1-D convolutional layer and the decoder is a 1-D deconvolutional layer (also known as transposed convolutional). In 
this scenario, the mixture weights replace the commonly used STFT representations. 

Since TasNet operates on waveform segments that can be as small as 5 ms, the system can be implemented in real-time with very low latency. In addition to having lower latency, TasNet outperforms the state-of-art STFT-based system. In applications that do not require real-time processing, a noncausal separation module can also be used to further improve the performance by using information from the entire signal.

\section{Model Description}
\label{sec:model}

\begin{figure*}[!ht]
\centering
\includegraphics[width=15cm]{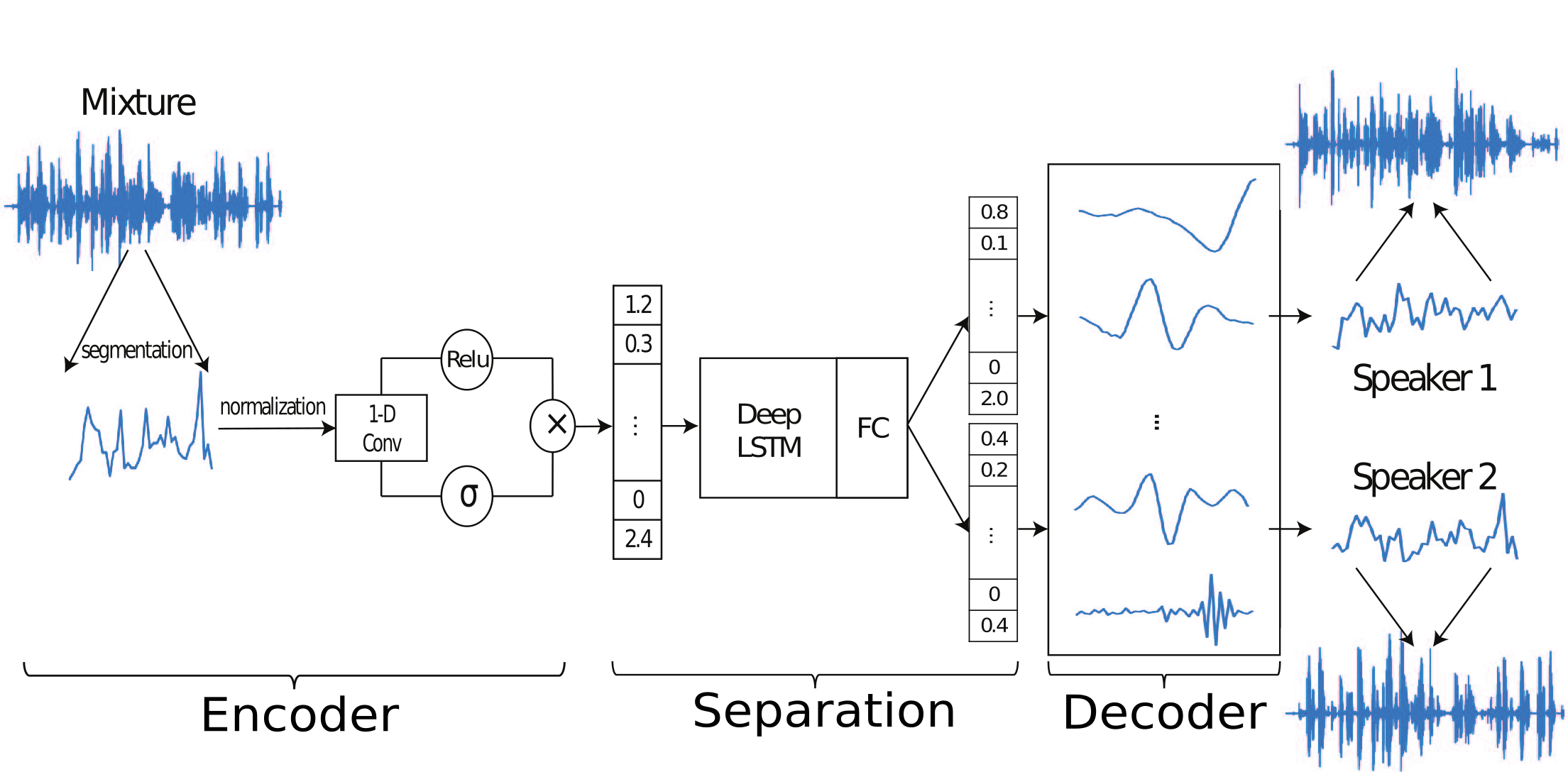}
\caption{Time-domain Audio Separation Network (TasNet) models the signal in the time-domain using encoder-decoder framework, and perform the source separation on nonnegative encoder outputs. Separation is achieved by estimating source masks that are applied to mixture weights to reconstruct the sources. The source weights are then synthesized by the decoder.}
\label{fig:flow}
\end{figure*}

\subsection{Problem formulation}

The problem of single-channel speech separation is formulated as estimating $C$ sources $s_1(t), \ldots, s_c(t)$, given the discrete waveform of the mixture $x(t)$
\begin{align}
x(t) = \sum_{i=1}^C s_i(t)
\label{eqn:prob}
\end{align}
We first segment the mixture and clean sources into $K$ non-overlapping vectors of length $L$ samples, $\vec{x}_k \in \mathbb{R}^{1\times L}$ (note that $K$ varies from utterance to utterance)
\begin{align}
\begin{cases}
\vec{x}_k = x(t)\\ \vec{s}_{i,k}=s_i(t)
\end{cases}
t\in [kL,(k+1)L),\, k = 1, 2, \dotsc, K
\end{align}
For simplicity, we drop the notation $k$ where there is no ambiguity. Each segment of mixture and clean signals can be represented by a nonnegative weighted sum of $N$ basis signals \vec{B} = [$\vec{b}_1, \vec{b}_2, \ldots, \vec{b}_N] \in \mathbb{R}^{N\times L}$
\begin{align}
\begin{cases}
\vec{x} = \vec{w} \vec{B}\\\\
\vec{s}_{i} = \vec{d}_{i} \vec{B}
\end{cases}
\label{eqn:ica}
\end{align}
where $\vec{w}\in \mathbb{R}^{1\times N}$ is the mixture weight vector, and $\vec{d}_{i}\in \mathbb{R}^{1\times N}$ is the weight vector for the source $i$. Separating the sources in this representation is then reformulated as estimating the weight matrix of each source $\vec{d}_i \in \mathbb{R}^{1\times N}$ given the mixture weight $\vec{w}$, subject to:
\begin{align}
\vec{w} = \sum_{i=1}^C \vec{d}_i
\label{eqn:weight}
\end{align}

Because all weights ($\vec{w}, \vec{d}_i$) are nonnegative, estimating the weight of each source can be thought of as finding its corresponding mask-like vector, $\vec{m}_i$, which is applied to the mixture weight, $\vec{w}$, to recover $\vec{D}_i$:
\begin{gather}
\vec{w} = \sum_{i=1}^C \vec{w} \odot (\vec{d}_i \oslash \vec{w}) := \vec{w} \odot \sum_{i=1}^C  \vec{m}_i\\
\vec{d}_i = \vec{m}_i \odot \vec{w}
\label{eqn:mask}
\end{gather}
where $\vec{m}_{i}\in \mathbb{R}^{1\times N}$ represents the relative contribution source $i$ to the mixture weight matrix, and $\odot$ and $\oslash$ denotes element-wise multiplication and division. 

In comparison to other matrix factorization algorithms such as ICA where the basis signals are required to have distinct statistical properties or explicit frequency band preferences, no such constraints are imposed here. Instead, the basis signals are jointly optimized with the other parameters of the separation network during training. Moreover, the synthesis of the source signal from the weights and basis signals is done directly in the time-domain, unlike the inverse STFT step which is needed in T-F based solutions. 

\subsection{Network design}
Figure~\ref{fig:flow} shows the structure of the network. It contains three parts: an encoder for estimating the mixture weight, a separation module, and a decoder for source waveform reconstruction. The combination of the encoder and the decoder modules construct a nonnegative autoencoder for the waveform of the mixture, where the nonnegative weights are calculated by the encoder and the basis signals are the 1-D filters in the decoder. The separation is performed on the mixture weight matrix using a subnetwork that estimates a mask for each source.

\subsubsection{Encoder for mixture weight calculation}
\label{sec:mix}

The estimation of the nonnegative mixture weight $\vec{w}_k$ for segment $k$ is done by a 1-D gated convolutional layer
\begin{align}
\vec{w}_k = ReLU(\vec{x}_k \circledast \vec{U}) \odot \sigma(\vec{x}_k \circledast \vec{V} ),\quad k = 1, 2, \dotsc, K
\label{eqn:conv}
\end{align}
where $\vec{U}\in \mathbb{R}^{N\times L}$ and $\vec{V}\in \mathbb{R}^{N \times L}$ are $N$ vectors with length $L$, and $\vec{w}_k\in \mathbb{R}^{1\times N}$ is the mixture weight vector. $\sigma$ denotes the Sigmoid activation function and $\circledast$ denotes convolution operator. $\vec{x}_k\in \mathbb{R}^{1\times L}$ is the $k$-th segment of the entire mixture signal $x(t)$ with length $L$, and is normalized to have unit $L^2$ norm to reduce the variability. The convolution is applied on the rows (time dimension).

This step is motivated by the gated CNN approach that is used in language modeling \cite{dauphin2017language}, and empirically it performs significantly better than using only ReLU or Sigmoid in our system. 

\subsubsection{Separation network}
\label{sec:source}
The estimation of the source masks is done with a deep LSTM network to model the time dependencies across the $K$ segments, followed by a fully-connected layer with Softmax activation function for mask generation. The input to the LSTM network is the sequence of $K$ mixture weight vectors $\vec{w}_1, \dotsc \vec{w}_K \in \mathbb{R}^{1\times N}$, and the output of the network for source $i$ is $K$ mask vectors $\vec{m}_{i,1}, \ldots, \vec{m}_{i,K} \in \mathbb{R}^{1\times N}$. The procedure for estimation of the masks is the same as the T-F mask estimation in \cite{kolbaek2017multitalker}, where a set of masks are generated by several LSTM layers followed by a fully-connected layer with Softmax function as activation.

To speed up and stabilize the training process, we normalize the mixture weight vector $\vec{w}_k$ in a way similar to layer normalization \cite{ba2016layer}
% well yes, then just layer-normalization is ok its just strange to have 3 hyphens can you say differently?
\begin{gather}
\hat{\vec{w}}_k = \frac{\vec{g}}{\sigma}\otimes (\vec{w}_k - \mu) + \vec{b}, \quad k = 1, 2, \dotsc, K \\
\mu =\frac{1}{N}\sum_{j=1}^N \vec{w}_{k,j} \quad \sigma = \sqrt{\frac{1}{N}\sum_{j=1}^N (\vec{w}_{k,j} - \mu)^2}
\label{eqn:ln}
\end{gather}
where parameters $\vec{g} \in \mathbb{R}^{1\times N}$ and $\vec{b} \in \mathbb{R}^{1\times N}$ are gain and bias vectors that are jointly optimized with the network. This normalization step results in scale invariant mixture weight vectors and also enables more efficient training of the LSTM layers.

% not sure if we still need this

%In addition to using $\vec{W}_k$ for estimating $\vec{M}_{i,k}$, a context window can be included to provide more information to the separation network. To add a one-segment context window, we concatenate $[\vec{W}_{k-1}, \vec{W}_k, \vec{W}_{k+1}]$ as the input to the deep LSTM network, and the output remains as $\vec{M}_{i,k}$. For first and last segments, a zero vector is appended accordingly. Note that adding the context window increases the minimum latency of the system from $L$ to $2L$.

Starting from the second LSTM layer, an identity skip connection \cite{he2016identity} is added between every two LSTM layers to enhance the gradient flow and accelerate the training process.

\subsubsection{Decoder for waveform reconstruction}
\label{sec:syn}

The separation network produces a mask matrix for each source $i$ $\vec{M}_i = [\vec{m}_{i,1}, \ldots, \vec{m}_{i,K}] \in \mathbb{R}^{K\times N}$ from the mixture weight $\hat{\vec{W}} = [\hat{\vec{w}}_1, \ldots, \hat{\vec{w}}_K] \in \mathbb{R}^{K\times N}$ across all the $K$ segments. The source weight matrices can then be calculated by
\begin{align}
\vec{D}_i = \vec{W} \odot \vec{M}_i
\label{eqn:source}
\end{align}
where $\vec{D}_i = [\vec{d}_{i,1}, \ldots, \vec{d}_{i,K}] \in \mathbb{R}^{K\times N}$ is the weight matrix for source $i$. Note that $\vec{M}_i$ is applied to the original mixture weight $\vec{W} = [\vec{w}_1, \ldots, \vec{w}_K]$ instead of normalized weight $\hat{\vec{W}}$. The time-domain synthesis of the sources is done by matrix multiplication between $\vec{D}_i$ and the basis signals $\vec{B}\in \mathbb{R}^{N\times L}$
\begin{align}
\vec{S}_{i} = \vec{D}_{i} \vec{B}
\label{eqn:syn}
\end{align}

For each segment, this operation can also be formulated as a linear deconvolutional operation (also known as transposed convolution) \cite{dumoulin2016guide}, where each row in $\vec{B}$ corresponds to a 1-D filter which is jointly learned together with the other parts of the network. This is the inverse operation of the convolutional layer in Section~\ref{sec:mix}.

Finally we scale the recovered signals to reverse the effect of $L^2$ normalization of $\vec{x}_k$ discussed in Section~\ref{sec:mix}. Concatenating the recoveries across all segments reconstruct the entire signal for each source.
\begin{align}
s_i(t) = [\vec{S}_{i,k}], \quad k = 1, 2, \dotsc, K
\end{align}

\subsubsection{Training objective}
\label{sec:obj}

Since the output of the network are the waveforms of the estimated clean signals, we can directly use source-to-distortion ratio (SDR) as our training target. Here we use scale-invariant source-to-noise ratio (SI-SNR), which is used as the evaluation metric in place of the standard SDR in \cite{isik2016single, chen2017deep}, as the training target. The SI-SNR is defined as:
\begin{gather}
    \vec{s}_{target} = \frac{\langle \hat{\vec{s}}, \vec{s} \rangle \vec{s}}{\left \| \vec{s} \right \|^2}\\
    \vec{e}_{noise} = \hat{\vec{s}} - \vec{s}_{target}\\
    \text{SI-SNR} := 10\,log_{10}\frac{\left \| \vec{s}_{target} \right \|^2}{\left \| \vec{e}_{noise} \right \|^2}
    \label{eqn:sdr}
\end{gather}
where $\hat{\vec{s}} \in \mathbb{R}^{1\times t}$ and $\vec{s}\in \mathbb{R}^{1\times t}$ are the estimated and target clean sources respectively, $t$ denotes the length of the signals, and $\hat{\vec{s}}$ and $\vec{s}$ are both normalized to have zero-mean to ensure scale-invariance. Permutation invariant training (PIT) \cite{kolbaek2017multitalker} is applied during training to remedy the source permutation problem \cite{isik2016single, kolbaek2017multitalker, chen2017deep}.

\section{Experiments}
\label{sec:exp}

\subsection{Dataset}

We evaluated our system on two-speaker speech separation problem using WSJ0-2mix dataset \cite{isik2016single, kolbaek2017multitalker, chen2017deep}, which contains 30 hours of training and 10 hours of validation data. The mixtures are generated by randomly selecting utterances from different speakers in Wall Street Journal (WSJ0) training set si\_tr\_s, and mixing them at random signal-to-noise ratios (SNR) between 0 dB and 5 dB. Five hours of evaluation set is generated in the same way, using utterances from 16 unseen speakers from si\_dt\_05 and si\_et\_05 in the WSJ0 dataset. To reduce the computational cost, the waveforms were down-sampled to 8 kHz.

\subsection{Network configuration}

The parameters of the system include the segment length $L$, the number of basis signals $N$, and the configuration of the deep LSTM separation network. Using a grid search, we found optimal $L$ to be 40 samples (5 ms at 8 kHz) and $N$ to be 500. We designed a 4 layer deep uni-directional LSTM network with 1000 hidden units in each layer, followed by a fully-connected layer with 1000 hidden units that generates two 500-dimensional mask vectors. For the noncausal configuration with bi-directional LSTM layers, the number of hidden units in each layer is set to 500 for each direction. An identical skip connection is added between the output of the second and last LSTM layers.

During training, the batch size is set to 128, and the initial learning rate is set to $3e^{-4}$ for the causal system (LSTM) and $1e^{-3}$ for the noncausal system (BLSTM). We halve the learning rate if the accuracy on validation set is not improved in 3 consecutive epochs. The criteria for early stopping is no decrease in the cost function on the validation set for 10 epochs. Adam \cite{kingma2014adam} is used as the optimization algorithm. No further regularization or training procedures were used.

We apply curriculum training strategy \cite{bengio2009curriculum} in a similar fashion with \cite{isik2016single, chen2017deep}. We start the training the network on 0.5 second long utterances, and continue training on 4 second long utterances afterward.

\subsection{Evaluation metrics}

For comparison with previous studies, we evaluated our system with both SI-SNR improvement (SI-SNRi) and SDR improvement (SDRi) metrics used in \cite{isik2016single, kolbaek2017multitalker, chen2017deep}, where the SI-SNR is defined in Section~\ref{sec:obj}, and the standard SDR is proposed in \cite{vincent2006performance}.

\subsection{Results and analysis}

Table~\ref{tab:result} shows the performance of our system as well as three state-of-art deep speech separation systems, Deep Clustering (DPCL++, \cite{isik2016single}), Permutation Invariant Training (PIT, \cite{kolbaek2017multitalker}), and Deep Attractor Network (DANet, \cite{chen2017deep}). Here TasNet-LSTM represents the causal configuration with uni-directional LSTM layers. TasNet-BLSTM corresponds to the system with bi-directional LSTM layers which is noncausal and cannot be implemented in real-time. For the other systems, we show the best performance reported on this dataset. 

We see that with causal configuration, the proposed TasNet system significantly outperforms the state-of-art causal system which uses a T-F representation as input. Under the noncausal configuration, our system outperforms all the previous systems, including the two-stage systems DPCL++ and uPIT-BLSTM-ST which have a second-stage enhancement network. Note that our system does not contain any regularizers such as recurrent dropout (DPCL++) or post-clustering steps for mask estimation (DANet).

Table~\ref{tab:speed} compares the latency of different causal systems. The latency of a system $T_{tot}$ is expressed in two parts: $T_i$ is the initial delay of the system that is required in order to receive enough samples to produce the first output. $T_p$ is the processing time for a segment, estimated as the average per-segment processing time across the entire test set. The model was pre-loaded on a Titan X Pascal GPU before the separation of the first segment started. The average processing speed per segment in our system is less than 0.23 ms, resulting in a total system latency of 5.23 ms. In comparison, a STFT-based system requires at least 32 ms time interval to start the processing, in addition to the processing time required for calculation of STFT, separation, and inverse STFT. This enables our system to preform in situation that can tolerate only short latency, such as hearing devices and telecommunication applications. 

\begin{table}[!t]
\centering
\caption{SI-SNR (dB) and SDR (dB) for different methods on WSJ0-2mix dataset.}
\vspace{0.2cm}
\label{tab:result}
\begin{tabular}{c|c|c|c}
\thline
Method & Causal & SI-SNRi & SDRi\\
\hline
uPIT-LSTM \cite{kolbaek2017multitalker} & \checkmark & -- & 7.0 \\
TasNet-LSTM & \checkmark & 7.7 & \bf{8.0} \\
%TasNet-LSTM-C & \checkmark & \bf{7.8} & \bf{8.1}\\
\hline
DPCL++ \cite{isik2016single} & \texttimes & \bf{10.8} & --  \\
DANet \cite{chen2017deep} & \texttimes & 10.5 & -- \\
uPIT-BLSTM-ST \cite{kolbaek2017multitalker} & \texttimes & -- & 10.0\\
TasNet-BLSTM & \texttimes & \bf{10.8} & \bf{11.1} \\
\thline
\end{tabular}
\end{table}

\begin{table}[!t]
\centering
\caption{Minimum latency (ms) of causal methods.}
\vspace{0.2cm}
\label{tab:speed}
\begin{tabular}{c|c|c|c}
\thline
Method & $T_i$ & $T_p$ & $T_{tot}$\\
\hline
uPIT-LSTM \cite{kolbaek2017multitalker} & 32 & -- & \textgreater32\\
TasNet-LSTM & 5 & 0.23 & \bf{5.23} \\
%TasNet-LSTM-C & 10 & 0.24 & 10.24 \\
\thline
\end{tabular}
\end{table}

\begin{figure}[!htp]

\begin{minipage}[b]{1.0\linewidth}
  \centering
  \centerline{\includegraphics[width=7.5cm]{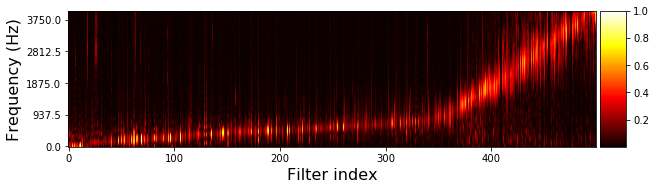}}
%  \vspace{1.5cm}
  \centerline{(a)}\medskip
\end{minipage}
\hfill
\begin{minipage}[b]{1.0\linewidth}
  \centering
  \centerline{\includegraphics[width=7.5cm]{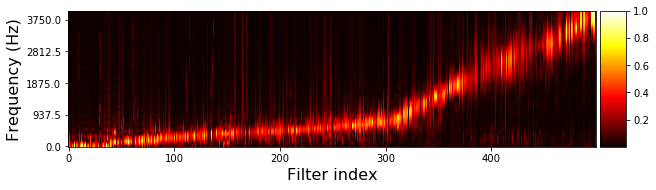}}
%  \vspace{1.5cm}
  \centerline{(b)}\medskip
\end{minipage}
\caption{Frequency response of basis signals in (a) causal and (b) noncausal networks.}
\label{fig:basis}
\end{figure}

To investigate the properties of the basis signals $\vec{B}$, we visualized the magnitude of their Fourier transform in both causal and noncausal networks. Figure~\ref{fig:basis} shows the frequency response of the basis signals sorted by their center frequencies (i.e. the bin index corresponding to the the peak magnitude). We observe a continuous transition from low to high frequency, showing that the system has learned to perform a spectral decomposition of the waveform, similar to the finding in \cite{sainath2015learning}. We also observe that the frequency bandwidth increases with center frequency similar to mel-filterbanks. In contrast, the basis signals in TasNet have a higher resolution in lower frequencies compared to Mel and STFT. In fact, 60\% of the basis signals have center frequencies below 1 kHz (Fig. \ref{fig:basis}), which may indicate the importance of low-frequency resolution for accurate speech separation. Further analysis of the network representation and transformation may lead to better understanding of how the network separates competing speakers \cite{nagamine2017understanding}.

\section{Conclusion}
\label{sec:conclude}

In this paper, we proposed a deep learning speech separation system that directly operates on the sound waveforms. Using an autoencoder framework, we represent the waveform as nonnegative weighted sum of a set of learned basis signals. The time-domain separation problem then is solved by estimating the source masks that are applied to the mixture weights. Experiments showed that our system was 6 times faster compared to the state-of-art STFT-based systems, and achieved significantly better speech separation performance. Audio samples are available at \cite{web2018tasnet}.

\section{Acknowledgement}
This work was funded by a grant from National Institute of Health, NIDCD, DC014279, National Science Foundation CAREER Award, and the Pew Charitable Trusts.
\vfill\pagebreak
\bibliographystyle{IEEEbib}
\bibliography{refs}

\end{document}